\documentclass[prb,twocolumn, floatfix]{revtex4}

\usepackage{graphicx}
\usepackage{bm, amsmath, amssymb}

\newcommand{\nbar}[0]{\bar{n}}

\newcommand{\qext}[0]{Q_\textrm{\small{ext}}}

\newcommand{\pin}[0]{P_\textrm{\small{in}}}

\begin{document}

\title{Single crystal silicon capacitors with low microwave loss in the single photon regime}
\author{S. J. Weber}
\author{K.W. Murch}
\author{D.H. Slichter}
\author{R. Vijay}
\author{I. Siddiqi}
\email[Corresponding author: ]{irfan@berkeley.edu}
\affiliation{Quantum Nanoelectronics Laboratory, Department of Physics, University of California, Berkeley CA 94720}

\date{\today}

\begin{abstract}

We have fabricated superconducting microwave resonators in a lumped element geometry using single crystal silicon dielectric parallel plate capacitors with C $>2$ pF.   Aluminum devices with resonant frequencies between 4.0 and 6.5 GHz exhibited an average internal quality factor $Q_i$ of $2\times10^5$ in the single photon excitation regime at T = 20 mK. Attributing the observed loss to the capacitive element, our measurements place and upper bound on the loss tangent of the silicon dielectric layer of $\tan \delta_i  = 5 \times 10^{-6}$. This level of loss is an order of magnitude lower than is currently observed in structures incorporating amorphous dielectric materials, thus making single crystal silicon capacitors an attractive, robust route for realizing long-lived quantum circuits. \end{abstract}

\maketitle

Superconducting circuits such as quantum bits\cite{mart05loss}, inductance based photon detectors, \cite{day03} and linear resonators require low loss reactive microwave components to achieve a high internal quality factor $Q_i$. At frequencies below the energy gap, superconducting metallic wires function as high quality inductors whose remnant loss is presumably due only to surface and interface defects and/or the presence of non-equilibrium quasiparticles.\cite{PhysRevB.72.014517,lena11} Capacitive elements, particularly structures involving amorphous dielectric layers, typically have significantly greater loss.\cite{ocon08amorphous} Currently, the highest $Q_i$ values are achieved in circuits comprised of planar interdigitated capacitors on single crystal silicon or sapphire substrates.  Such circuits achieve $Q_i$ values in the $2\times10^5$ range with an average resonator population of one photon,\cite{khal10} but it is difficult to realize capacitances larger than about 1 pF in this geometry.   Higher capacitances can be obtained in a parallel plate geometry using either deposited amorphous dielectrics\cite{ocon08amorphous} such as SiO$_2$, $a$-Si:H, or SiN$_x$\cite{paik10SiNx} or vacuum gaps.\cite{cica10gap}  However, deposited dielectrics exhibit significant loss in the low temperature, low power regime due to the presence of two-level state (TLS) defects, \cite{mart05loss,shni05tls} limiting their utility for quantum circuits.   Current vacuum gap capacitors suffer from loss from necessary support structures and surface oxides, and are on par with the best deposited dielectrics, with a loss tangent $\tan \delta_i =1/Q_i= 2.2$-$3 \times 10^{-5}$ in the 4-8 GHz band.\cite{ocon08amorphous,paik10SiNx,cica10gap}   Crystalline dielectrics, such as silicon can exhibit low intrinsic loss and have been used in the fabrication of superconducting bolometers using a wafer bonding process. \cite{deni09}

In this letter, we report on the performance of linear microwave lumped element resonators incorporating parallel plate capacitors with a single crystal silicon dielectric layer.   At millikelvin temperatures and excitations corresponding to an average resonator occupation $\bar{n}$ of one photon, the typical operating conditions of superconducting qubits, we find that these capacitors exhibit lower loss in the 4.0-6.5 GHz range than state-of-the-art amorphous dielectric and vacuum gap capacitors, and are comparable to the best planar circuits on silicon.  Thus, they can potentially boost coherence times in quantum circuits currently employing amorphous dielectric materials, such as the phase qubit. Furthermore, these high quality, high capacitance elements can be used to directly investigate the intrinsic loss of Josephson junctions at GHz frequencies.   

 \begin{figure}
\includegraphics[angle = 0, width = 0.5\textwidth]{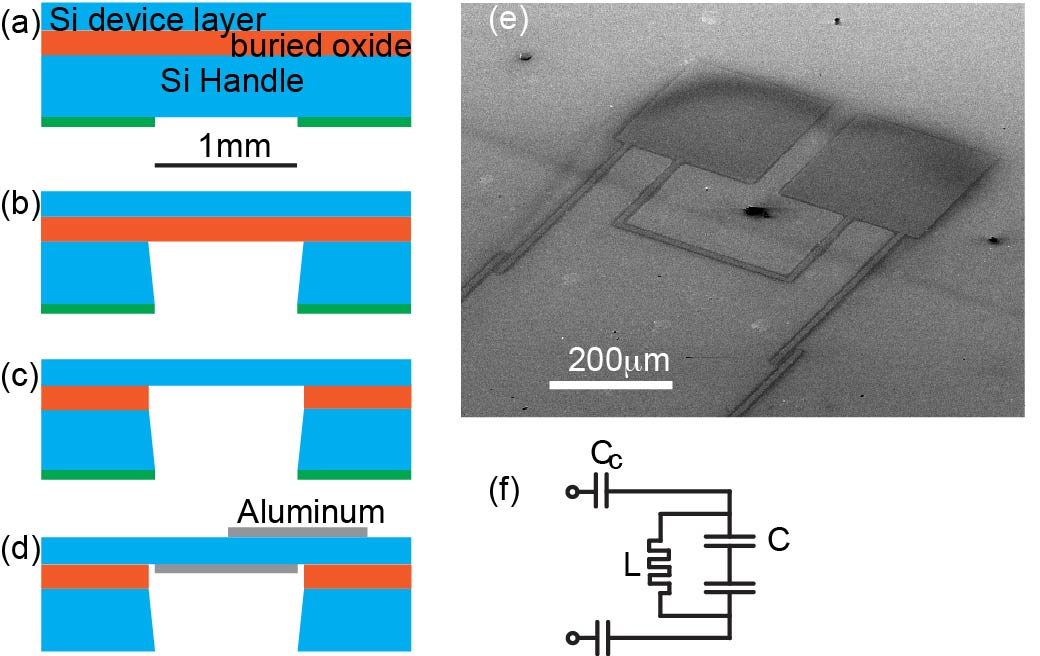}
\caption{\label{fig1} Parallel plate capacitors from single crystal silicon dielectric. The bottom of a silicon-on-insulator (SOI) wafer was patterned with resist (green) and etched using a deep silicon etch process (a,b). The buried oxide layer was removed and the top and bottom surfaces were patterned with aluminum (c,d).   (e) Scanning electron micrograph of the top surface of the resonator.   (f) Circuit model of the resonator with balanced excitation.}
\end{figure}

The fabrication and geometry of our lumped element resonators is detailed in Figure 1.  The devices consist of two superconducting metal layers separated by a $d=2$ $\mu$m thick dielectric layer of high resistivity silicon. Two pads on the top layer are capacitively coupled to the bottom metal layer, forming two parallel plate capacitors in series.  This structure is shunted with a $L\simeq 600$ pH superconducting meander inductor. There is no metal on the bottom surface in the vicinity of the inductor.  The resonator is isolated from the 50 $\Omega$ environment by coupling capacitors which set the external quality factor $\qext \simeq 10^5$.

The Si dielectric layer was formed from a commercially available (Ultrasil) high resistivity ($\rho> 1\textrm{ k}\Omega\textrm{-cm}$) silicon-on-insulator (SOI) wafer.  The handle wafer was patterned with 8 microns of SPR 220-7 resist, and etched using a deep silicon etch process, with alternating cycles of SF$_6$ plasma etching and C$_4$F$_8$ passivation. The buried oxide layer acted as the etch stop.  Both the buried oxide layer and the native oxide on top of the silicon device layer were removed by HF vapor. Ground planes were deposited on the bottom of the device layer using 300 nm of e-beam evaporated aluminum.  The top device layer was optically patterned and coated with 100 nm of aluminum.

Resonators were mounted in copper sample boxes inside superconducting and Cryoperm magnetic shields on the mixing chamber stage of a dilution refrigerator and cooled to 20 mK.   The samples were isolated from thermal microwave photons by 72 dB of attenuation on the input line and 3 cryogenic isolators preceding the cryogenic semiconductor amplifier on the output line.    The resonant response was obtained using a vector network analyzer.  The excitation power used to probe the resonators varied between $P_\textrm{\small{in}} =-94$ and $P_\textrm{\small{in}}=-157$ dBm, and for the one port devices we consider is related to the average photon number on resonance as $\nbar=4 P_\textrm{\small{in}} Q^2/(\qext \hbar \omega_0^2)$, where $\omega_0$ is the resonant frequency and the total quality factor $Q=( 1/Q_i + 1/\qext)^{-1}$.  The quality factors $Q_i$ and $\qext$ are determined by fitting to the resonance curve.  

\begin{figure}
\includegraphics[angle = 0, width = 0.5\textwidth]{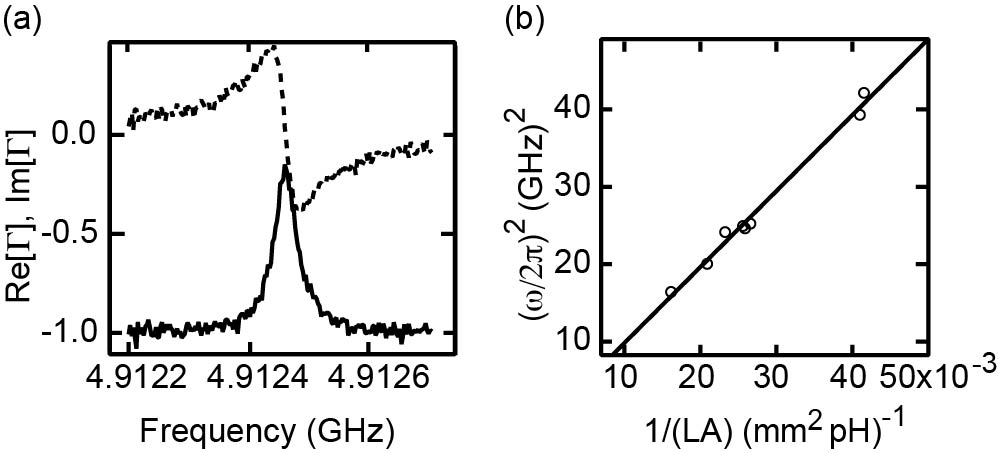}
\caption{\label{fig2} (a) The scaled and rotated real (solid) and imaginary (dashed) parts of the reflected signal from an LC resonator measured at 20 mK and $\pin  = -153$ dBm, corresponding to $\nbar=0.9$.  (b) The resonance frequency scales as $\omega_0^2 = 2/(L c A)$, where $A$ is the area of one of the capacitor pads; the solid line shows the expected scaling given independently derived values of $c$ and $L$. }
\end{figure}

Figure 2a shows the real and imaginary parts of the reflected microwave signal for a characteristic sample, measured at 20 mK with an average photon occupation of $\nbar=1$.  The device exhibits no spurious resonances in the 4-8 GHz band.  This was also confirmed by finite element simulations of the actual device geometry. In Figure 2b, we plot the squared resonance frequency of all of the samples tested.  Assuming an idealized parallel plate geometry, the resonance frequency is expected to scale as  $\omega_0^2 = 2/(LcA)$, where $A$ is the area of one of the capacitor pads, $L$ is the shunting inductance, and $c=\epsilon_r\epsilon_0/d = 51.4$ pF/mm$^2$ is the specific capacitance of the device layer, assuming a bulk value of $\epsilon_r=11.68$ for the dielectric layer.  Using values of the inductance obtained from finite element simulations, we plot the expected resonance frequency in Figure 2b as a solid black line. The slight deviations from theory are likely due to variations in the thickness of the device layer, which we measured to vary by as much as $20\%$ across the wafer.

\begin{table}[htdp]
\caption{Summary of aluminum resonator results.  The devices marked with an asterisk (*) did not have the native surface oxide removed before patterning.  The capacitance, $C$, of one of the parallel capacitors represents twice the total circuit capacitance due to the series geometry.}
\begin{center}
\begin{tabular}{l c c c ccc }
      \hline
\hline
$\omega_0/(2\pi)$(GHz)& &$Q_i(\nbar=1)$ && $\qext$& &$C$ (pF)\\
\hline
4.99223&$\quad$ &$2.4(2)\times 10^5$&$\quad$&$7.1\times 10^4$&$\quad$&3.25\\
6.26773&  &$1.5(2)\times 10^5 $&&$8.9\times 10^3$&&2.08\\
4.05102&  &$2.2(1)\times 10^5 $&&$2.1\times 10^5$&&4.68\\
4.91246&  &$1.8(1)\times 10^5 $&&$2.6\times 10^5$&&3.25\\
6.49231&  &$1.11(3)\times 10^5 $&&$9.1\times10^4$&&2.08\\
5.02214$^*$&		&$2.8(6)\times10^4 $&&$3.5\times 10^5$&&3.25\\
4.47342$^*$&		&$2.3(1)\times 10^4  $&&$4.5\times 10^5$&&4.68\\
4.96406$^*$&		&$3.27(5)\times 10^4 $&&$1.0\times 10^5$&&3.25\\
\hline
\hline
\end{tabular}
\end{center}
\label{default}
\end{table}%

   \begin{figure}
\includegraphics[angle = 0, width = 0.5\textwidth]{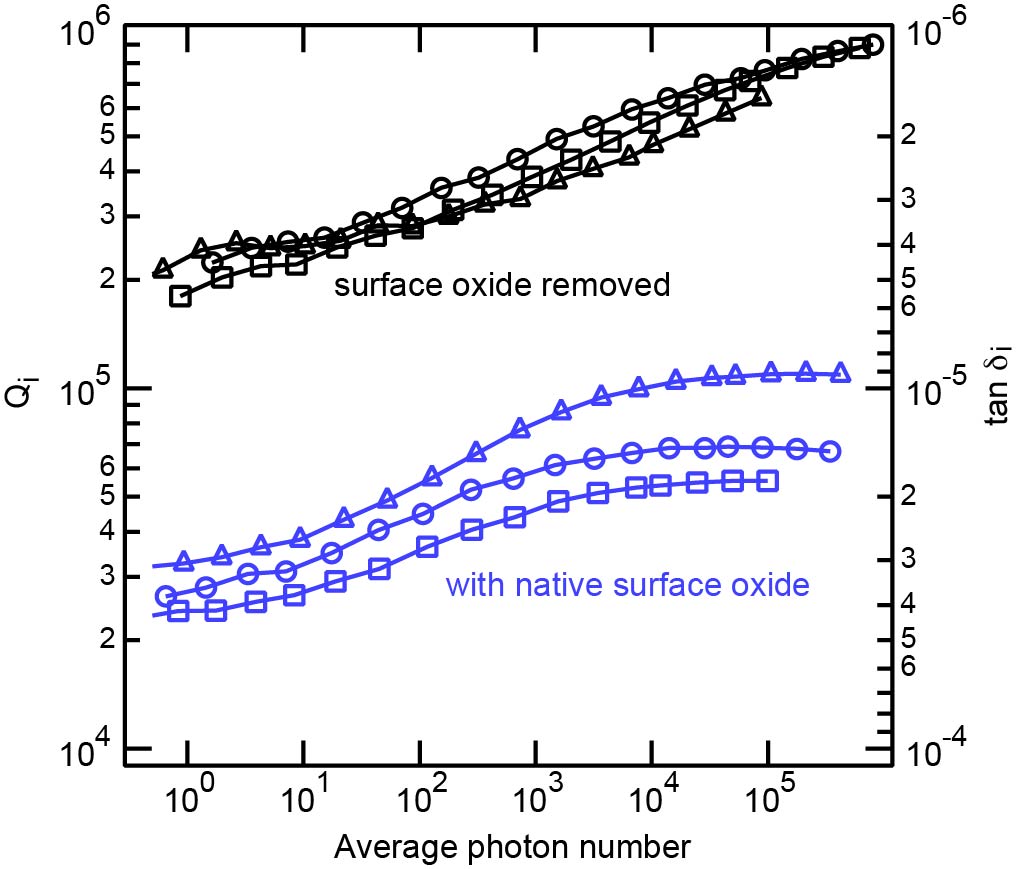}
\caption{\label{fig3} Internal quality factor of  aluminum resonators between 4 and 5 GHz.  Devices which had the native oxide stripped from the top of the device layer (black) exhibited lower losses  than devices with the native surface oxide present (blue).  Black symbols $\{\bigtriangleup,\circ, \Box \}$, correspond to $ \omega_0/(2\pi) =\{4.99223, 4.05102, 4.91246\}$ GHz resonators respectively.  Blue symbols correspond to $\omega_0/(2\pi) = \{4.96406,5.02214,4.47342\}$ GHz resonators.}
\end{figure}

In Figure 3, we plot the extracted internal quality factor versus excitation power expressed in terms of average photon number for aluminum lumped element resonators in the 4 to 5 GHz range.  Table 1 summarizes the results of all of the lumped element resonators we tested.  We note that stripping the native oxide off the top of the device layer prior to patterning the surface increased the resonator Q nearly tenfold. Of the devices tested with this surface treatment, the average intrinsic loss tangent in the low temperature, single average photon regime was $\tan \delta_i = 5.4\times 10^{-6}$.  If we assume that the observed losses are solely due to the dielectric layer, our measurements are consistent with the current reported microwave loss tangent of single crystal silicon under these experimental conditions.\cite{ocon08amorphous}

Our results indicate that single crystal silicon capacitors are very promising for use in high quality factor superconducting circuits which require $>$ pF of shunting capacitance, such as the phase qubit. Furthermore, incorporating these elements into classical non-linear resonators consisting of a capacitively shunted Josephson junction will allow precise microwave characterization of the high frequency loss and $1/f$ noise in weak-link\cite{vija10} and tunnel junctions. \cite{vanh04}

In conclusion, we have developed a technique to fabricate lumped element parallel plate capacitors with losses that are nearly an order of magnitude lower than existing technologies.  Our technique is simple, robust, uses commercially available materials and standard processing techniques.

This research was funded by the Office of the Director of National Intelligence (ODNI), Intelligence Advanced Research Projects Activity (IARPA), through the Army Research Office. All statements of fact, opinion or conclusions contained herein are those of the authors and should not be construed as representing the official views or policies of IARPA, the ODNI, or the US Government.  D.H.S. acknowledges support from a Hertz Foundation Fellowship endowed by Big George Ventures.  R.V. acknowledges funding from AFOSR under Grant No. FA9550-08-1-0104

\pagebreak


\end{document}